\documentclass[%
 reprint,
 amsmath,amssymb,
 aps,
]{revtex4-1}

\usepackage{graphicx}
\usepackage{dcolumn}
\usepackage{bm}
\usepackage[mathlines]{lineno}
\usepackage{physics}
\usepackage{braket}
\usepackage{caption}
\usepackage{subcaption}

\begin{document}

\preprint{APS/123-QED}

\title{Efficient quantum memory for heralded single photons generated by cavity-enhanced spontaneous parametric downconversion}

\author{Yu-Chih Tseng$^{1}$}
\author{Yan-Cheng Wei$^{1,2}$}
\author{Ying-Cheng Chen$^{1,3}$}

\affiliation{$^{1}$Institute of Atomic and Molecular Sciences, Academia Sinica, Taipei 10617, Taiwan}

\affiliation{$^{2}$Department of Physics, National Taiwan University, Taipei 10617, Taiwan}

\affiliation{$^{3}$Center for Quantum Technology, Hsinchu 30013, Taiwan}

\date{\today}

\begin{abstract}
We interface a spontaneous parametric down conversion (SPDC) crystal and a cold atomic ensemble and demonstrate a highly efficient quantum memory through polarization-encoded single-photon qubits. Specifically, narrowband heralded single photons from a cavity-enhanced SPDC source is stored using cold atomic ensemble, with $\sim 70 \%$ storage-and-retrieval efficiency and $\sim 10 \mu s$ storage time at $50\%$ efficiency. To prevent the degradation after storage, we also manipulate the single-photon wave profile so that the retrieved non-classical nature of single photon is preserved. On the other hand, the dual-rail storage is used for storing polarization-encoded qubits, and the corrected fidelity of flying qubits after storage reaches $\sim 97 \%$. The results pave the way toward large-scale quantum network.
\end{abstract}

\maketitle




\section{Introduction}

Quantum networks relies on efficient transfer between flying qubits and stationary quantum nodes \cite{Kimble2008, Cao2020}. To facilitate the distribution of entanglement over large-scale network, the synchronization between two distant nodes are essential for many protocols, such as quantum repeaters \cite{RevModPhys.83.33}, linear optics quantum computation (LOQC) \cite{Knill2001}. To do that, quantum memories, devices that can store and retrieve flying qubits on demand, are critical tools. Thus, intensive efforts have been made on optical storage with promising results, such as $92\%$ storage efficiency (SE) with electromagnetically induced transparency (EIT) \cite{PhysRevLett.120.183602}, $30\%$ with off resonance
Raman interaction \cite{Reim2010}, $87\%$ with photon echo technique \cite{Hedges2010}, and $8.4\%$ with Autler-Townes-splitting protocol (ATS) \cite{Saglamyurek2018}. However, most of them are working on attenuated or single-photon-level coherent light.  It should be pointed out that although some protocols can be realised through single-photon-level classical light source, developing single-photon Fock state is essential for a wider range of applications on quantum information processing, such as LOQC \cite{Knill2001} and entanglement transfer\cite{PhysRevLett.80.2245, Cao2020}. While fewer works have addressed quantum storage for single photons, pushing toward universal quantum networks remains challenging.

To generate single photon, the possible platforms include single atoms inside a cavity \cite{McKeever1992}, four-wave
mixing from the cold atomic ensemble \cite{PhysRevLett.100.183603}, NV (nitrogen-vacancy) centers in diamond \cite{5192205, PhysRevLett.108.143601}, the Duan–Lukin–Cirac–Zoller
protocol (DLCZ) by cold atomic ensembles \cite{Duan2001, Cao2020}. The merit of cold-atom-based single photon source is that the central frequencies match the atomic transitions, and photons are inherently narrow-band, which enables the stronger interaction and yields higher efficiency. Up to date, works toward highly efficient quantum storage of cold-atom-generated single photon have been demonstrated with  SE $85 \%$, the fidelity$> 99\%$\cite{Wang2019}, and with SE $> 85 \%$ and preserving antibunching nature \cite{Cao2020}. Yet, the tedious setup for atom-based photon source is relatively arduous to scale up, which is difficult for more sophisticated manipulation on flying qubits.

\begin{figure*}
    \centering
    \includegraphics[width = 18cm]{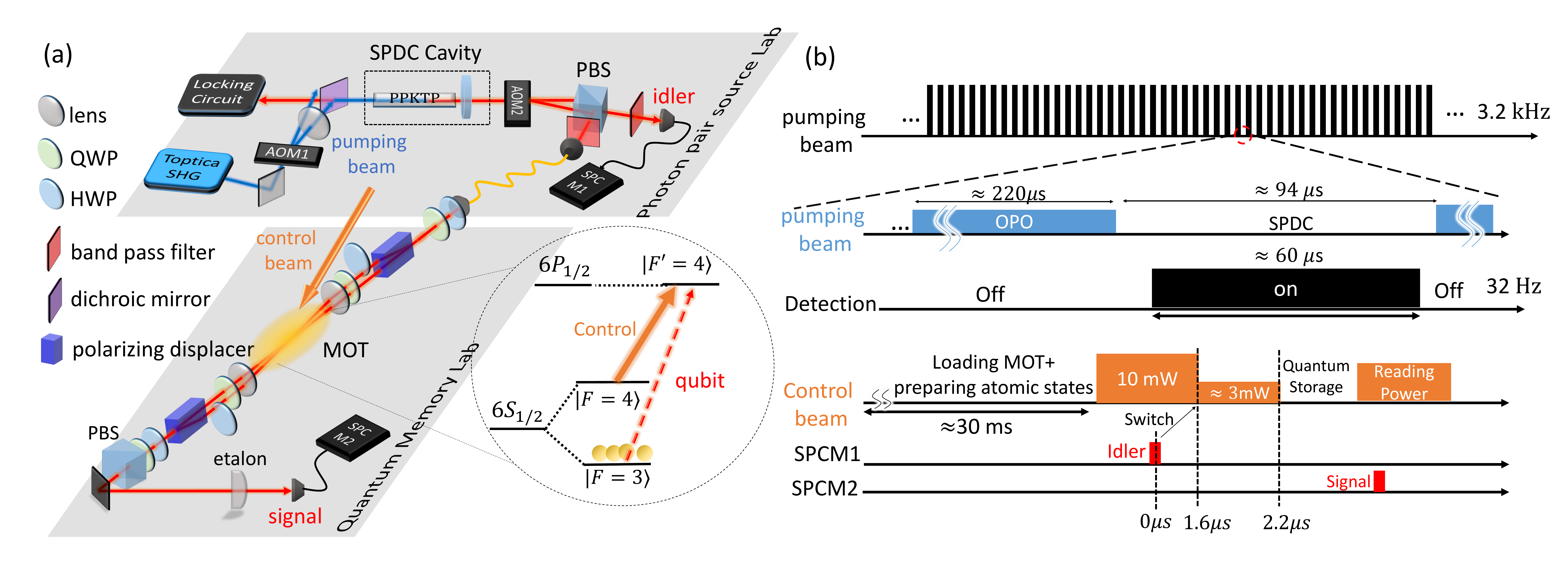}
    \caption{(a) The experimental setup: cavity-enhanced SPDC photon source and EIT-based  cold-atom quantum memory with dual-rail polarization qubits, shown on the upper and lower part, respectively. The inset is the energy-level diagram of Cs $D_1$ transitions. Signal photons as polarization-encoded qubits drive the $\ket{F=3}$ to $\ket{F'=4}$ transition. The strong control field drives $\ket{F=4}$ to $\ket{F'=4}$ transition. (b) Timing sequence of the experiment. The pumping power switches between high-power and low-power phases at $3.2 kHz$, while the cold-atom system operates at a 32 Hz repetition rate. After triggered by idler photons, the control beam power switches through write-store-retrieve sequence.}
    \label{fig:setup}
\end{figure*}

Conversely, spontaneous parametric down conversion (SPDC) crystals, one of the most widely used photonic entanglement sources, are suitable for large-scale protocols \cite{PhysRevResearch.2.033155}. The drawbacks of them are ultra-broad bandwiths ($\sim $THz), hindering the direct interaction with atoms. The issue can be solved elegantly through placing crystals into high-fitness optical cavities, suppressing the spectral linewidths of single photons to $\sim$MHz level without sacrificing photons \cite{Tsai_2018, Zhang2011}. Nevertheless, although the quantum storage of SPDC-based photon sources have been realize \cite{PhysRevResearch.2.033155, Zhang2011, Akiba2009, Clausen2011, Seri:18, PhysRevLett.123.080502, PhysRevLett.112.040504}, the SEs are typically low. The best SE is $36 \%$ \cite{PhysRevResearch.2.033155} in previous works, which has not yet beat $50\%$ the no-cloning theorem limits \cite{PhysRevA.64.010301} and loss tolerance in cluster state computation\cite{PhysRevLett.97.120501}. The major difficulties include noises pollution \cite{Wang2019} (see Appendix \ref{app: noises for g2}), mode stabilization of SPDC crystals \cite{Tsai_2018}. Many efforts have been made to generate bright, narrowband \cite{Tsai_2018}, single-mode single photon sources \cite{PhysRevA.83.061803}, in order to maximize the light-matter interaction and thus increase SE.

In this work, we demonstrate a highly efficient quantum memory with $\sim 70\%$ SE of SPDC photons, which to our knowledge is the best record for crystal-generated photon sources. As for quantum memories, the choice of cold atomic ensemble enables us to reach high storage efficiency and long storage times \cite{PhysRevLett.120.183602}. We aim to build a compact photon source for easily scaling up, while maintaining the efficient transfer between channels and a node. Such a setup allows us to conduct advanced manipulations on photonic states with large-scale photon sources, such as post-selected entanglements \cite{PhysRevLett.101.190501, PhysRevLett.102.063603}, and the generated photons can be efficiently stored and delivered using a cold atomic quantum memory. The crystal-atom-interfaced system forms a quantum node in a quantum network with more freedom to manipulate photonic states.

\section{Experimental Setup}
The experimental setup consists of a photon-pair source and a cold atom system, as shown in Fig.\ref{fig:setup}a. The nondegenerate, narrowband photon pairs are generated from a cavity-enhanced SPDC setup with a single longitudinal mode. A Toptica frequency-doubling laser, which serves as the pump beam with a wavelength of 447 nm, pumps the type-II periodically poled KTiOPO4 (PPKTP) crystal in the high-finesse linear cavity. Due to the similar reflectivity of the mirrors on the two sides of the linear cavity, the down-converted photons travel nearly 50/50 forward and backward. During the experiment, the pump power alternates between a high-power ($\sim 18 mW$) phase and a low-power $\sim 50 \mu W$ phase with a switching rate of 3.2 kHz. $70\%$ duty cycle operates with the high-power phase, which is controlled by an acousto-optic modulator (AOM1) (see Fig.\ref{fig:setup} b). During the high-power phase (which is corresponding to the optical parametric oscillator, OPO), the backward beam provides the signal for cavity frequency stabilization. The forward beam is modulated by AOM2, which serves as an optical switch with the same driving frequency as AOM1. AOM2 is out of phase to AOM1 with a $20\%$ duty cycle. This allows the beam to pass only during the low-power phase (SPDC) and prevents the photon detectors (SPCM) from being damaged. Signal and idler photons are separated by a polarizing beam splitter (PBS) and are further sent to the cold-atom system and received as triggers, respectively. The linewidth of the heralded single photons is 2.2 MHz. The OPO output light is beating with a reference laser, of which the frequency is controlled by a double-passed AOM. The central frequency of the signal beam is locked at $^{133}Cs$ $D_1$ transition (895 nm) by locking the beat frequency to a certain value. More technical details can be referred to \cite{Tsai_2018}.

A magneto-optical trap (MOT) of cesium with  cigar-shaped atomic clouds is used to implement the quantum memory with electromagnetically induced transparency (EIT) protocol \cite{Tsai_2018, PhysRevLett.120.183602}. The EIT memory is operated at $D_1$ line, with the signal photons drive the $\ket{F=3}$ to $\ket{F'=4}$ transition and the strong control field drives the $\ket{F=4}$ to $\ket{F'=4}$ transition, as shown in the sub-graph of Fig.\ref{fig:setup} a. The choice of $D_1$ line allows us to reduce the control-intensity-dependent ground-state decoherence rate due to the off-resonant excitation of the control field to the nearby transition \cite{PhysRevLett.120.183602, Tsai_2018}. The temporally dark and compressed
MOT and  Zeeman-state optical pumping are used to increase the optical depth \cite{PhysRevA.90.055401}. The achieved optical depth is $\approx \{235 , 306, 334\}$ when a MOT repetition rate set at $\{32, 16, 8\}$Hz, respectively. The optical depth is determined by the EIT spectral fitting by setting the decay rate of the optical coherence ($\gamma_{ge}$) to be $\sim 0.7\Gamma$, where $\Gamma = 2\pi \times 4.56$ MHz is the linewidth of $D_1$ transition. This decay rate is larger than an ideal value of 0.5$\Gamma$ (solely due to the spontaneous decay) because the contribution due to laser linewidth and laser frequency fluctuation is considered\cite{PhysRevLett.120.183602}.

The idler photons are detected by SPCM1 for heralding while signal photons are sent to the quantum memory laboratory through a $400$m optical fiber, which is used to induce a $\approx 2 \mu$s temporal delay. This optical delay is technically essential such that the optical switch for the control beam can response before the signal photons passing through the MOT cell. Then, the signal photons pass through the polarization displacer to map the polarizations of the qubit into the two spatial modes \cite{VernazGris2018, Wang2019}. A half-wave plate is added in one of the spatial mode, such that both modes have the same polarization before interacting with cold atomic ensembles. Because the atomic population is optically pumped towards the rightmost Zeeman states to increase the optical depth, such a dual-rail setup allows us to store any polarization of flying qubits without lossing much of optical depth\cite{VernazGris2018, Wang2019}. Before coming into the MOT cell, the signal beam is focused by a lens to an intensity $e^{-2}$ diameter of $\approx 90 \mu m$ around the atomic clouds while the control beam is collimated with a diameter of $\approx 520 \mu m$. The angle between the signal and control beam is $\approx 4^o$. This angle is carefully selected such that the noises from the control beam are minimized while the ground state decoherence rate is still suppressed at $\approx 5 \times 10^{-3} \Gamma$, which majorly comes from the residual Doppler broadening. After passing through the MOT cell, the signal beam passes a reverse setup as that before entering into the MOT cell, such that both spatial modes are collimated and combined. The signal photons are then collected by SPCM2 before passing four irises, two elatons (Quantaser FPE001), and a bandpass filter (Semrock FF01-900/25), which are all used to filter out the noises. 
The reduction of external noises is essential to preserve the quantum nature of the single photons, as discussed in Appendix.\ref{app: noise model}. The overall collection efficiency, defined as the power ratio between that at the photon counter and the output right after SPDC, is $\approx 2.8 \%$.

Fig.\ref{fig:setup}b depicts the time sequence of our experiment. The repetition rates for the photon pair source and MOT are 3.2 kHz and 32 (or 16) Hz, respectively. Within each MOT period, there is a $\approx 60 ms$ for cold atoms loading and states preparation and then the MOT is turned off for $0.6$ ms. During the MOT off time, the photon-pair system is at the photon-pair mode for $\approx 94 \mu s$. Waiting for $34 \mu$s to allow the leaking photons from the OPO phase to a negligible level\cite{PhysRevResearch.2.033155}, the latter $\approx 60 \mu s$ time window is left for collecting data. Once the SPCM1 detects an idler photon, the electronic signal is sent to the optical switch, and the control beam is switched to an optimal intensity ($3$ mW) for quantum storage \cite{PhysRevLett.120.183602}. Before that, the control beam is set to a stronger intensity ($10$ mW) in order to clear out atomic population at the $\ket{F=4}$ state (see Appendix.\ref{app: noises for g2}). After the signal photons entering into the atomic ensemble, the control beam is ramp off to convert the signal photons into atomic spin-waves. After storing for a given time, the control beam is ramp up to retrieve the signal photons. The single photons are then collected by SPCM2 and record by an oscilloscope (RTO2014).

\section{results}
\subsection{Single-Photon Source}
\begin{figure}[t]
    \centering
    \includegraphics[width = 8.5 cm]{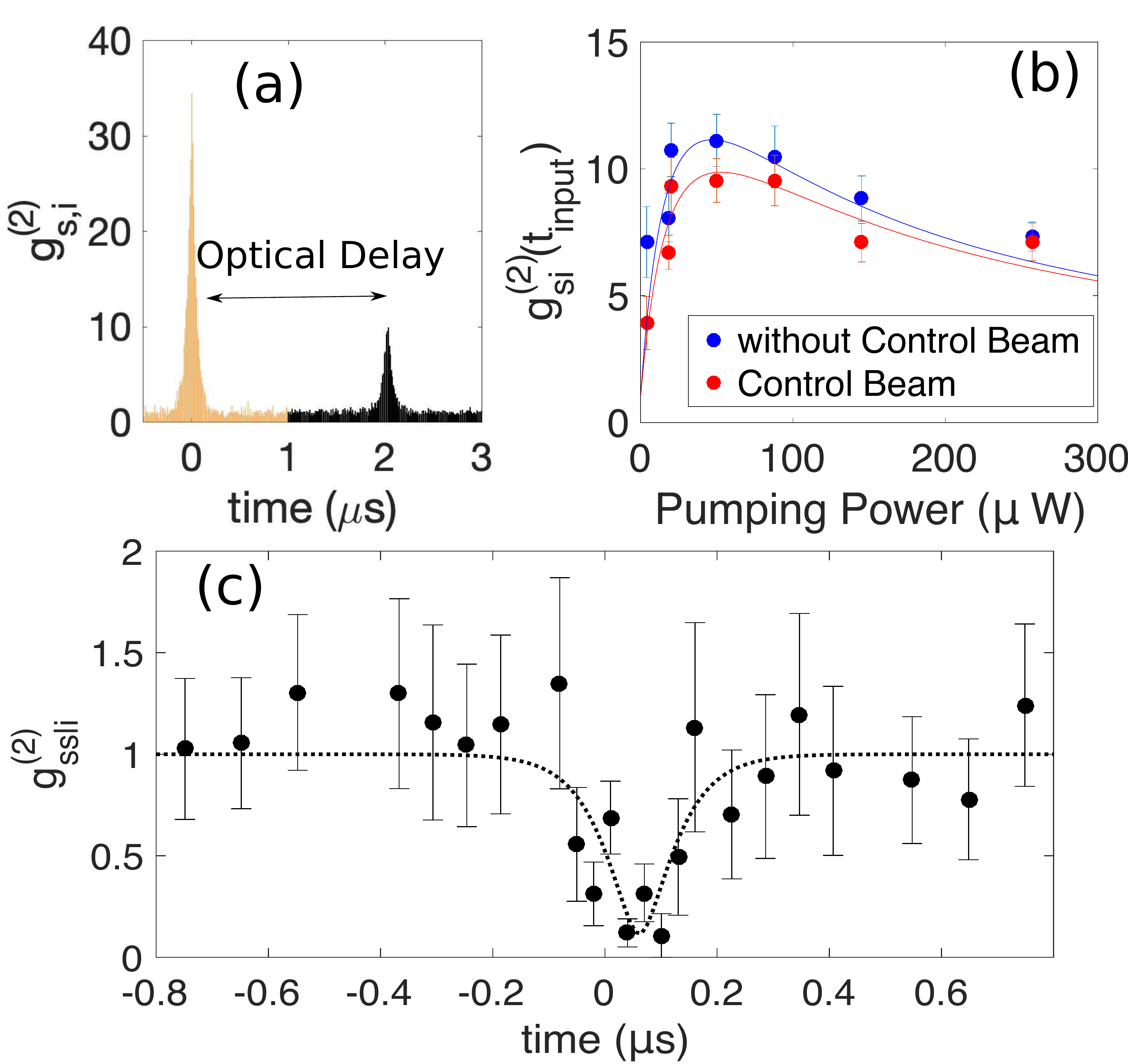}
    \caption{(a) The black-(yellow-)color points are measured normalized cross-correlation functions of biphoton pairs after arriving the quantum memory setup (right after single photon setup). (b) The peak $g^{(2)}_{si}$ of photon pairs measured after arriving at the quantum memory laboratory. The solid curves are the fitting curves by Eq. \ref{g2} with $\kappa^2 = A \times (\text{Pumping Power})$. The fitting parameters are $\{\Gamma_{cav}, n_s, n_i, A\} = \{22.9409 , 0.4561, 0.2134, 0.0680\}$ for the case with control beam and $\{22.9409 , 0.3515, 0.2134, 0.0680\}$ for the case without control beam. (c) Normalized autocorrelation function of photon pairs measured after arriving at the quantum memory laboratory}
    \label{fig:sys parms}
\end{figure}
We first characterize the nonclassical property of the heralded single photons by measuring the normalized cross-correlation function $g^{(2)}_{si}(t;\tau) = G^{(2)}_{si}(t;\tau)/(G^{(1)}_{s}(t) G^{(1)}_{i}(t+\tau))$, where $G^{(2)}_{si}(t;\tau) = \langle{\hat{a}^\dagger_i(t) \hat{a}^\dagger_s(t + \tau) \hat{a}_s(t + \tau) \hat{a}_i(t)} \rangle{}$ and $G^{(1)}_{s(i)}(t) = \langle{\hat{a}^\dagger_{s(i)}(t) \hat{a}_{s(i)}(t) }\rangle{}$.  In the experiment, the normalized cross-correlation function can be determined by
$g^{(2)}_{si} = p_{s,i}/(p_s p_i)$, where $p_{s} (p_i)$ is the probability of detecting the signal (idler) photons, and $p_{s,i}$ is the probability of detecting the coincidence events of both signal and idler photons\cite{PhysRevResearch.2.033155}. $g^{(2)}$ is determined by the bin width $=10$. If one assumes the second-order auto-correlations of signal and idler
$1 \leq [g_{ss}, g_{ii}] \leq 2$, $g^{(2)}_{si} > 2$  manifests the  violation of classical property \cite{Tsai_2018, Clausen2011, Kuzmich_2003}. Thus, it is desirable to pursue a better $g^{(2)}_{si}$ for enlarging the violation factor \cite{Tsai_2018}. Therefore, we scan through the pump power to determine the best working point, as shown in Fig.\ref{fig:sys parms}b. The detailed theory used to model the data can be referred to Appendix.\ref{app: noise model}. At the best working point, the maximum $g^{(2)}_{si}$ are $\approx 35$ and $\approx 10$ right after the photon source and after arriving at quantum memory setup, respectively, as shown in Fig.\ref{fig:sys parms}a.

The anti-bunching behavior of the single-photon nature is also confirmed by measuring the conditional second order autocorrelation function for the signal photons $g^{(2)}_{ss|i}$, which should be $0$ for an ideal single-photon Fock state and $1$ for coherent light. Our measured $g^{(2)}_{ss|i} $ is $0.199(0.069)$ after arriving at quantum memory setup, as shown in Fig.\ref{fig:sys parms}c. More details can be referred to Appendix.\ref{app: auto}

\begin{figure}[t!]
    \centering
    \includegraphics[width = 9.cm]{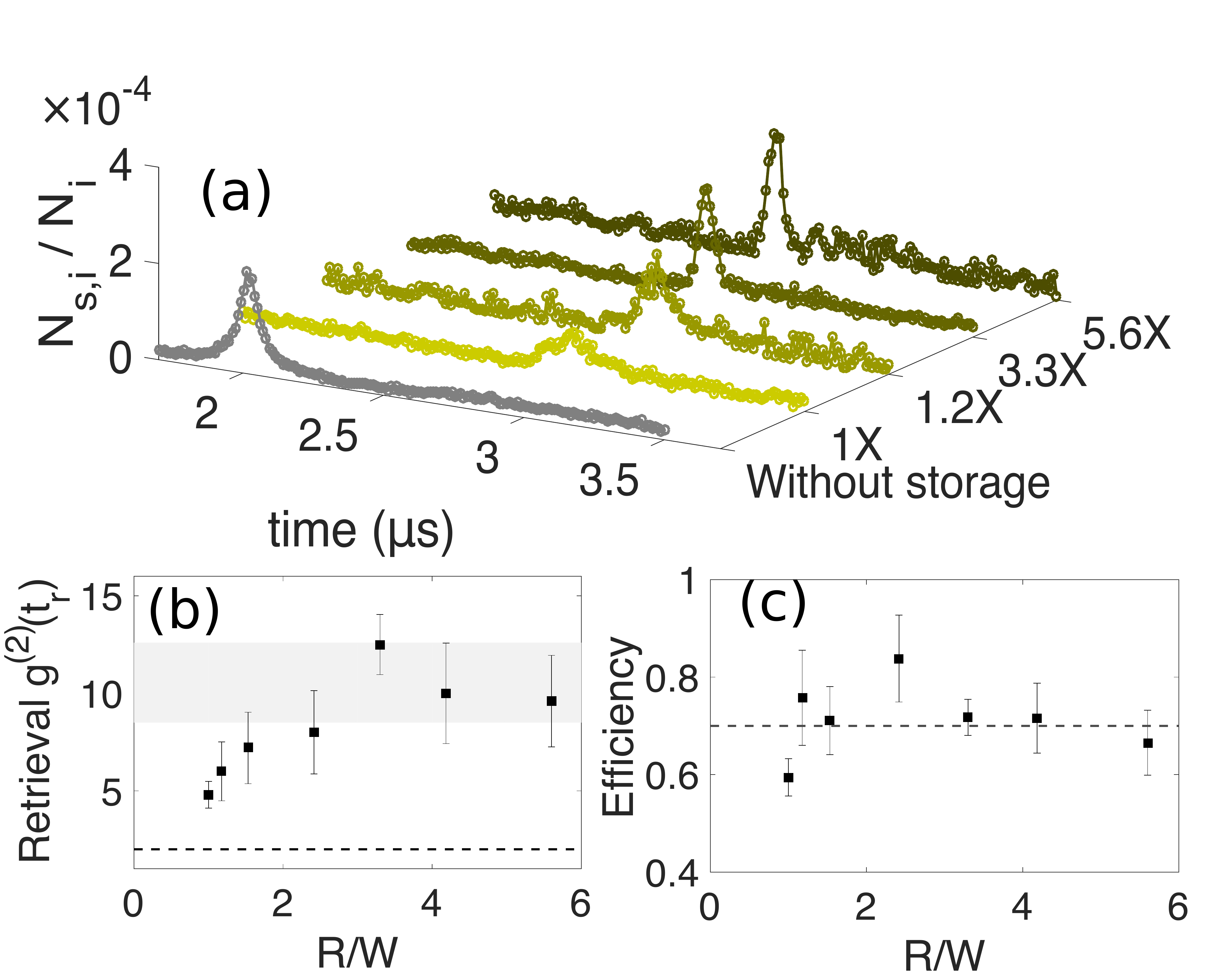}
    \caption{The notations $R:$ reading power of the control beam; $W:$ writing power of the control beam; $N_{s,i}$ is the coincidence counts of signal and idler photons; $N_i$ is the idler photon count. (a) Raw data of the input single-photon waveform and the retrieved waveform with $R/W = \{1, 1.2, 3.3, 5.6\}$, respectively.
    (b) The peak $g^{(2)}_{si}$ of the retrieved photons versus the $R/W$. $t_r$ denotes the retrieval time. The gray area denotes the range of $g^{(2)}_{si}$ peak for the input signal photons. The dashed curve denotes the classical limit ($g^{(2)}_{si}=2$). (c) The efficiency of the retrieved photons versus the $R/W$. The dashed line denotes an efficiency of 0.7. }
    \label{fig:diff RW}
\end{figure}

\subsection{Storage of Heralded Single Photons}
We then conduct the experiment on storage of heralded single photons. The repetition rate of the experiment is set at $16$Hz. We vary the control intensity during the retrieval process to manipulate the waveform and non-classical correlation of the retrieved signal photons. The data with a storage time of $400$ns for various ratios between the reading to writing control intensity are shown in Fig.\ref{fig:diff RW}a. It is evident that the peak height of the retrieved signal pulse is higher for a stronger reading control field. 
We emphasize that the implementation of EIT memory in
$D_1$ line allows us to increase $g^{(2)}_{si}$ of signal photon without sacrificing the storage efficiency, compared to the case of $D_2$ 
transition. This is due to the significant reduction of the off-resonant coupling of the control field to the nearby transition for the $D_1$ line as compared to the $D_2$\cite{PhysRevResearch.2.033155, PhysRevLett.120.183602}. The peak $g^{(2)}_{si}$ and storage-and-retrieval efficiency (SE) versus the $R/W$ control intensity ratio (from 1-5.6) are depicted in Fig.\ref{fig:diff RW}b and c, respectively. The details of the determination of the efficiency are described in Appendix.\ref{app: SE definition}. The efficiencies (SE) are around $\sim 70\%$ for different $R/W$ control intensity ratio. A slight deviation from the average efficiency for some data points may due to the drift of the system parameters (e. g. optical depth) since it takes $6\sim8$ hours for collecting each data set. 

Although the peak height of the signal pulse is greater for a stronger $R/W$ control intensity ratio as shown in Fig.\ref{fig:diff RW}a, the peak $g^{(2)}_{si}$ for the retrieved signal photons still saturate at higher intensity ratios as shown in Fig.\ref{fig:diff RW}b. The peak $g^{(2)}_{si}$ increases by a factor of $\sim 2$ for $R/W$ ratio from 1 to 3.3. This means that one can manipulate the waveform (or bandwidth) and the non-classical correlation of the retrieved signal photons by the EIT memory through varying the writing control intensity\cite{PhysRevResearch.2.033155}. The saturation of the peak $g^{(2)}_{si}$ at high writing control intensity is due to the increasing in the noise background which may come from the control leakage and Raman-induced noises. More discussions on the noises are shown in Appendix.\ref{app: noises for g2}. In all cases, the peak $g^{(2)}_{si}$ are all larger than 2. Therefore, the non-classical property of the heralded single photons is preserved by the EIT memory. The highest retrieved $g^{(2)}_{si}$ peak reaches $12.54 (1.27)$ at $R/W = 3.3$. 

\begin{figure}[t!]
    \centering
    \includegraphics[width = 8.3cm]{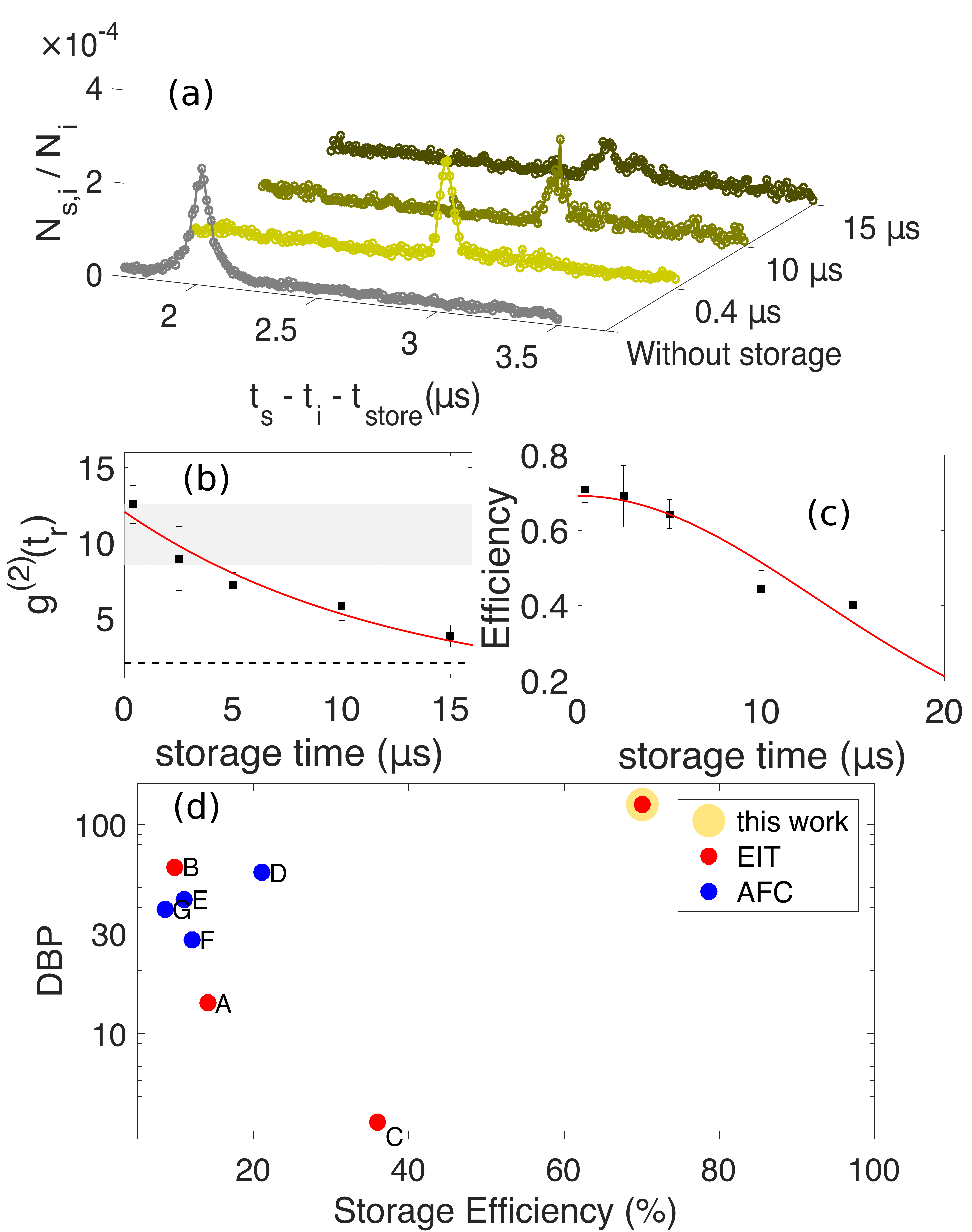}
    \caption{(a) The raw data of the retrieved signal photons for various storage times of $0.4 \mu s, 10\mu s, 15\mu s $. $t_{store}$: storage time; $t_s$: signal photon arrival time; $t_i$: idler photon arrival time. (b) The $g^{(2)}_{si}$ of the retrieved pulses (black squares) versus the storage time and the fitted curve (red solid line). The fitted curve is $g^{(2)}_{si,0} e^{-t/\tau}$ \cite{PhysRevLett.120.183602}, where $g^{(2)}_{si,0} = 12.1$ and $\tau = 12.1 \mu s$. The dashed line of $g^{(2)}_{si} \leq 2$ is the classical bound, below which is the classical regime. (c) The efficiency (black squares) versus the storage time and the fitted curve (red solid line). The fitted curve is $\text{SE}_0 e^{-t^2/\tau^2}$ \cite{PhysRevLett.120.183602}, where $\text{SE}_0 = 0.69$ and $\tau = 18.4 \mu s$.  (d) The comparison figure of the delay-bandwidth-product (DBP) versus the efficiency for the up-to-date works on quantum storage of heralded single photons generated by the SPDC source. A:\cite{Akiba2009}, B:\cite{Zhang2011}, C: \cite{PhysRevResearch.2.033155}; D:\cite{Clausen2011};E:\cite{PhysRevLett.123.080502}; F:\cite{Seri:18}; G:\cite{PhysRevLett.112.040504}}.
    \label{fig:storage}
\end{figure}

Next, we measure the retrieved signal photons for various storage time, as shown in Fig.\ref{fig:storage} a. The peak $g^{(2)}_{si}$ and efficiency versus the storage time are analyzed in Fig.\ref{fig:storage} b and c, respectively. The efficiency is below $50\%$ at $\approx 10 \mu s$, while the $g^{(2)}_{si}$ drops to half of the value without storage at $\approx 12.1 \mu s$. 
The delay-bandwidth product (DBP), usually defined as the ratio of the storage time when efficiency is 50\% to the FWHM duration of the input signal pulses, is an important figure of merit of the memory to quantify the ability to store the quantum nature. However, our work is the only one so far with an efficiency of larger than 50 \% for storage of single photons generated by the SPDC source. In order to compare with other works, we adopt the definition of DBP as the product of the storage time when retrieved $g^{(2)}_{si}$ is 50\% of its initial value with null storage time and the photon bandwidth. The comparison is shown in Fig.\ref{fig:storage} d\cite{PhysRevLett.120.183602, PhysRevLett.110.083601, wei2020broadband}. Although there is still a lot of room to improve our result, the achieved DBP of $\approx 138$ and the efficiency of 70\% are the best among these works.

\subsection{Storage of Polarization Qubits}

\begin{figure}[t!]
    \centering
    \includegraphics[width = 7.5 cm]{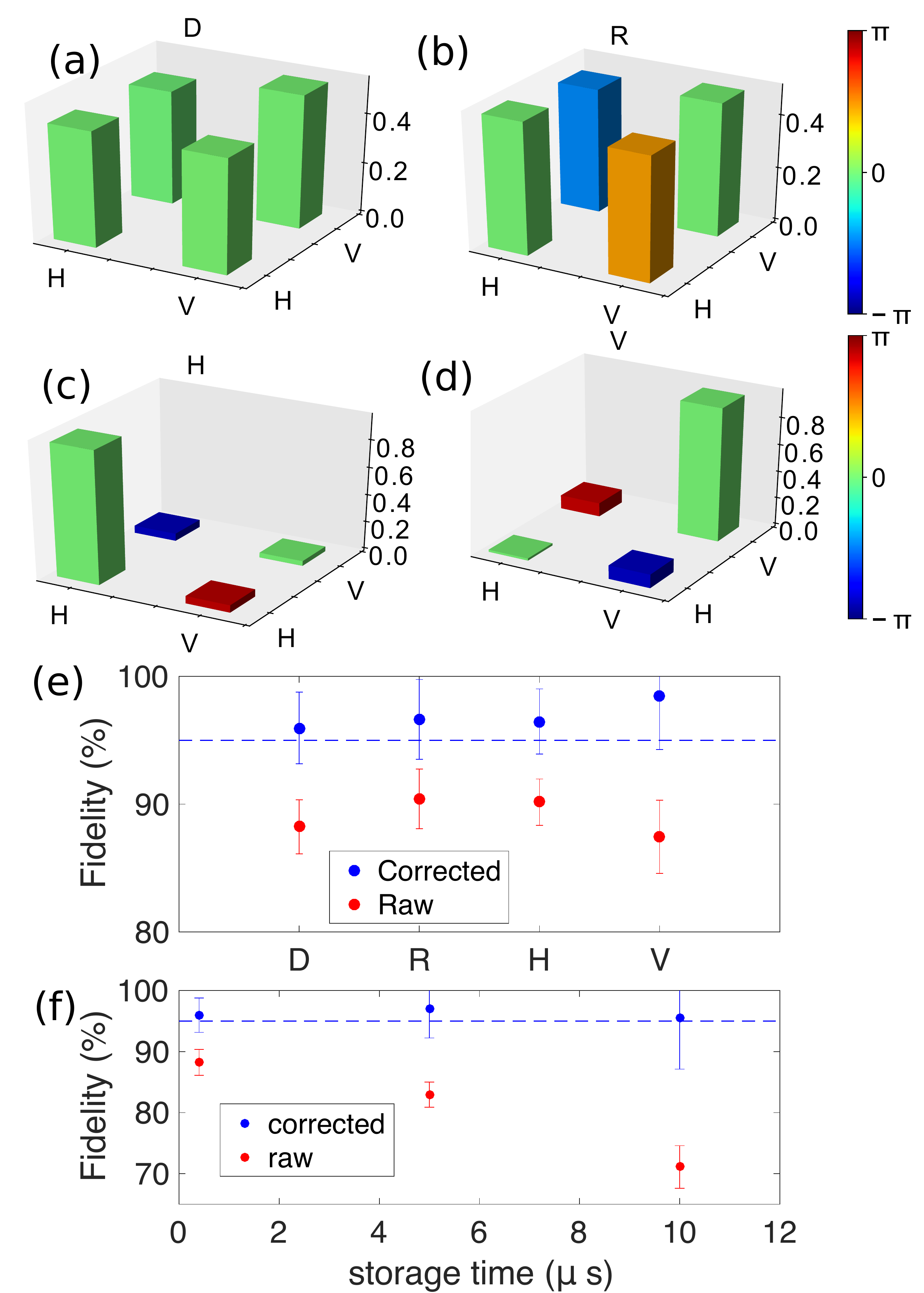}
    \caption{(a)-(d)The reconstructed density matrices of the retrieved signal photons with the input photon state prepared as $\ket{D}, \ket{R}, \ket{H},$ and $\ket{V}$, respectively. The colored bar on the right side indicates the phase of the components. The raw fidelity of $\ket{D}, \ket{R}, \ket{H}, \ket{V}$ are
    $88.23(2.12) \%$, $90.42(2.34) \%$, $90.16(1.81) \%$, $87.44(2.86) \%$. The corrected fidelity of $\ket{D}, \ket{R}, \ket{H}, \ket{V}$ are $95.96(2.81) \%$, $96.64(3.12) \%$, $96.47( 2.55) \%$, $98.45(4.17) \%$. The corrected fidelity is estimated by deducting the average background noise. (f) The fidelity of $\ket{D}$ versus the storage time. The raw fidelity drops with the declining fraction of the photon signal, while the corrected fidelity remains at the same level but with a larger uncertainty. The dashed lines in (c, d) indicate the $95\%$ fidelity.}
    \label{fig:Fid}
\end{figure}

A quantum memory should be able to store arbitrary states of qubit. We encode the polarization qubit into the heralded single photons by the dual-rail scheme\cite{VernazGris2018,Wang2019}, as shown in Fig.\ref{fig:setup}a. The polarization of the signal photons is mapped into the spatial mode. For example, for a qubit with a quantum state $\frac{1}{\sqrt{2}}(\ket{1}_H + e^{i \phi}\ket{1}_V)$, the dual-rail setup converts it into
\begin{equation}
    \frac{1}{\sqrt{2}}(\ket{1}_H + e^{i \phi}\ket{1}_V) \rightarrow  \frac{1}{\sqrt{2}}(\ket{1}_a + e^{i \phi}\ket{1}_b)
\end{equation}
where $H, V$ denote horizontal and vertical polarization; $a, b$ denote two different spatial modes (see Fig.\ref{fig:setup}a). The reversed setup is used after the MOT cell for mapping the spatial modes back into the original polarization states. We then use a set of wave-plates and polarization beam splitters to perform the quantum state tomography\cite{PhysRevA.64.052312}. The repetition rate of the experiment is set to $32$ Hz to speed up the data collection. 

Figures \ref{fig:Fid}a-d depict the reconstructed density matrices of the retrieved signal photons, of which the initial polarization states are $\ket{D}, \ket{R}, \ket{H},$ and $\ket{V}$, respectively. The average fidelity reaches $89.06(2.31)\%$. The performance is mainly limited by the low heralding efficiency ($\sim 4\times 10^{-3}$) and the background noises. Note that our mean photon number is $\approx 3.80 (0.04) \times 10^{-3}$ per heralded event, so the requirement on noises is very demanding \cite{VernazGris2018}. By subtracting the background noises, the average corrected fidelity is $96.88(3.22)\%$ (see Fig.\ref{fig:Fid}). The average storage efficiency is $55.76 (3.79)\%$. Compared with the single-rail setup, the possible reasons for the reduced efficiency are described below. First, the optical depth is reduced from $306$ to $235$ due to operating the experiment at a higher repetition rate (from 16 to 32 Hz), which is not favorable for a high efficiency. Second, the two spatial modes of the signal photons have a certain separation ($2.7$mm) such that the experienced optical depth may be reduced. In addition, the size of the control beam is not large enough to cover the two signal modes around its peak intensity region. The in-homogeneity of the control Rabi frequency may effectively induce a reduction in the signal transmission efficiency. 

We also repeat the above experiment for longer storage times, as depicted in Fig.\ref{fig:Fid}. The raw fidelity goes down as the storage time increases, while the corrected fidelity stays at nearly the same level but the uncertainty rises. The reason for the reduced raw fidelity is twofold. First, the memory efficiency drops when the storage time increases due to the decay of the spin waves resulting from the finite ground-state decoherence rate and the atomic motion. Second, the Raman-induced noise increases for longer storage time. This noise is due to the excitation of the atoms in the $|F=4\rangle$ ground state by the control field when it is turned back on for retrieval. The atoms in the $|F=4\rangle$ ground state may come from the cold atoms after the collisional relaxation or the background hot atoms in that state travelling through the control beam region when it is turned back on. We found that the later reason should be the major contribution, see Appendix.\ref{app: noises for g2} for more details. 

Finally, it is should be noted that the continuous-wave pumped cavity-enhanced SPDC source possesses strong spectral correlation between the signal and idler photons. This deteriorates the performance under the multi-source operation due to the low purity in joint frequency space \cite{PhysRevA.77.022312}, which can be resolved by exploiting the pulsed pumping scheme\cite{Zhang2011}.

\section{conclusion}
In summary, we demonstrate the quantum storage of heralded single photons and polarization qubits generated from the cavity-enhanced SPDC source. The  non-classical correlation is preserved and can be manipulated by the EIT memories. Both the achieved efficiency and delay-bandwidth product are the highest to date among the experiments using single photons generated from the SPDC source. The raw fidelity for the polarization qubits is around $89.06 (2.31)\%$ and the corrected fidelity is around $96.88 (3.22)\%$. The raw fidelity is limited by the low heralding efficiency (due to the optical losses) and the noises, which could be improved with more technical efforts. Our work paves the way towards large-scale quantum network.

\section{Supplementary materials}
\subsection{Measurement of autocorrelation function\label{app: auto}}
In the measurement, we use a beam splitter to split the original signal-photon channel into two and collect photons in these two arms. The conditional second-order autocorrelation function reads,
\begin{equation}
    g_{ss|i}^{(2)} = \frac{N_{s_1, s_2, i} N_i}{N_{s_1, i} N_{s_2, i}}
\end{equation}
, where $N_{s_1, s_2, i}$ is the three-fold coincidence counts of heralded idler photons and signal photons at both arms. $N_{s_{1(2)}, i}$ is the two-fold coincidence
counts of heralded idler photons and signal photons at one of two arms. $N_i$ is the idler photon counts \cite{Wang2019}.

\subsection{Determination of the efficiency \label{app: SE definition}}

The background of the coincidence count contains both the external noises (such as the stray photons or the leakage photons from the control beam) and the intrinsic noises (such as the leakage of the OPO photons into the SPDC phase or the accidental photons from the neighborhood photon pairs). The coincidence counts with and without storage are first subtracted by its corresponding average background count, determined by a time window away from the biphoton profile. The left coincidence counts with a profile of nearly two-sided exponential decay curve are those due to the true two-photon events. The total counts within a time window, centered around the peak count, of three times FWHM of the biphoton profile are calculated for both the data with and without storage. \cite{PhysRevA.83.061803, Tsai_2018}. The ratio of the counts with storage to that without storage is the efficiency.

\begin{figure}[t]
{\centering\includegraphics[width=8cm]{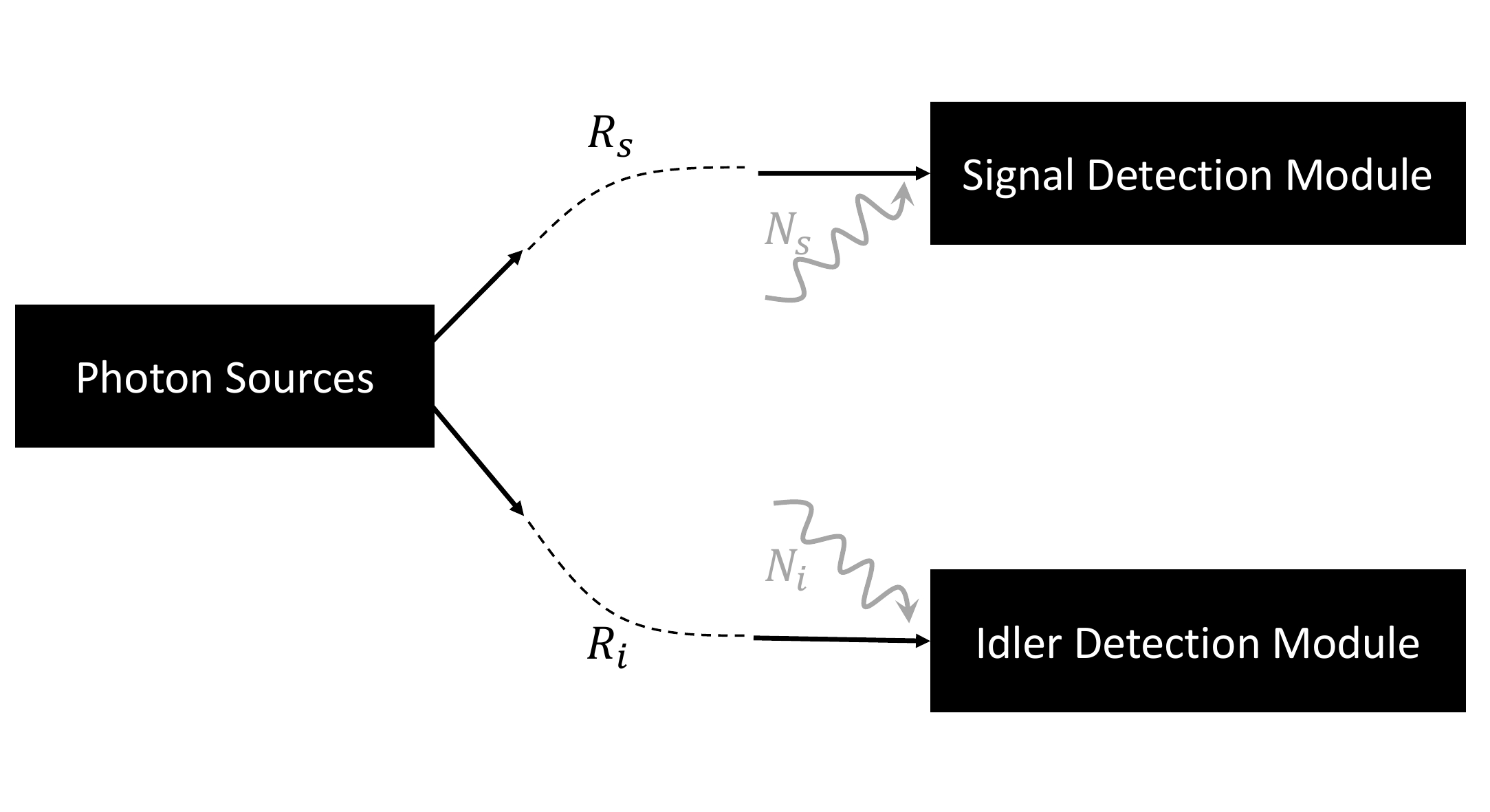}\\}
\caption{The illustration of the simple noise model discussed in Appendix.\ref{app: noise model}}
\label{NsModel}
\end{figure}

\subsection{Noise model for the photon pairs\label{app: noise model}}
The noise model for the photon pair detection is illustrated in Fig.\ref{NsModel}. The photon pairs are generated through a "photon source" box and sent to the two detection modules. There are inevitable losses along the propagation channel. We denote the total arrival efficiency for the signal (idler) photons as $R_{s (i)}$. In addition, both channels may suffer from noises. We denote the number of noise counts that entering the two channels as $N_i$ and $N_s$, respectively.

We here consider the cavity-enhanced SPDC photon-pair source, of which details can be referred to \cite{PhysRevA.83.061803}. Consider the cavity decay rates for signal and idler ($\Gamma_{s,cav} ,\Gamma_{i,cav}$) are nearly symmetry, $\Gamma_{s,cav} = \Gamma_{i,cav}  \equiv \Gamma_{cav}$, and the total decaying channel of signal and idler fields ($\gamma
_s, \gamma_i$) follow $\gamma_s = \gamma_i = \Gamma_{cav}$, which means we exclude other possible loss channels and only consider the coupling rate of the cavity. Based on the property of cavity-enhanced SPDC\cite{PhysRevA.83.061803},

\begin{equation}
    \begin{aligned}
    \bra{Vac} a^\dagger_{i}(t_i)a^\dagger_{s}(t_{s}) a_{s}(t_{s})  a_{i}(t_i)\ket{Vac}\\
    =(\kappa^2\times e^{-\Gamma_{cav} |t_s - t_i|} + \frac{4}{\Gamma_{cav}^2}\kappa^4) R_i R_s 
    \end{aligned}
\end{equation}
, where $\ket{Vac}$ is the vacuum state. The generation rate is
\begin{equation}
    \bra{Vac} a^\dagger_{s}(t) a_{s}(t) \ket{Vac} = \ \frac{2}{\Gamma_{cav}} \kappa^2 R_s
\end{equation}
\begin{equation}
    \bra{Vac} a^\dagger_{i}(t) a_{i}(t) \ket{Vac} = \ \frac{2}{\Gamma_{cav}} \kappa^2 R_i
\end{equation}
The coincidence count is  (notice it has used the commutation relationship to drop some terms)
\begin{equation}
    \begin{aligned}
    G^{(2)}_{si}(t_{s}, t_{i}) =& \bra{Vac} (a^\dagger_i(t_i)+ a^\dagger_{N_i}(t_i) )(a^\dagger_{s}(t_{s})+ \\& a^\dagger_{N_s}(t_{s})) (a_{s}(t_{s}) + a_{N_s}(t_{s}))  (a_i(t_i)+ a_{N_i}(t_{i}))\ket{Vac}\\
    =& \bra{Vac} a^\dagger_i(t_i)a^\dagger_{s}(t_{s}) a_{s}(t_{s})  a_i(t_i)\ket{Vac} + 
    \\
    & \bra{Vac} a^\dagger_{N_i}(t_i)a^\dagger_{s}(t_{s}) a_{s}(t_{s})  a_{N_i}(t_i)\ket{Vac} +
    \\
    & \bra{Vac} a^\dagger_{i}(t_i)a^\dagger_{N_s}(t_{s}) a_{N_s}(t_{s})  a_{i}(t_i)\ket{Vac} +
    \\
    & \bra{Vac} a^\dagger_{N_i}(t_i)a^\dagger_{N_s}(t_{s}) a_{N_s}(t_{s})  a_{N_i}(t_i)\ket{Vac} 
    \end{aligned}
\end{equation}

Here, the noises operators follow
\begin{equation}
    \begin{aligned}
    \bra{Vac}a^\dagger_{N_i}(t_i) a_{N_i}(t_i)\ket{Vac} = N_i\\
    \bra{Vac}a^\dagger_{N_s}(t_s) a_{N_s}(t_s)\ket{Vac} = N_s
    \end{aligned}
\end{equation}
Also, the noise should not have any correlation with other channels, meaning the occurrence of noises field commutes from other fields,
\begin{equation}
    \begin{aligned}
    & \bra{Vac} a^\dagger_{N_i}(t_i)a^\dagger_{N_s}(t_{s}) a_{N_s}(t_{s})  a_{N_i}(t_i)\ket{Vac} =\\
    & \bra{Vac} a^\dagger_{N_i}(t_i) a_{N_i}(t_i) a^\dagger_{N_s}(t_{s}) a_{N_s}(t_{s})  \ket{Vac}\\
    =& N_i N_s\\
   & \bra{Vac} a^\dagger_{N_i}(t_i)a^\dagger_{s}(t_{s}) a_{s}(t_{s})  a_{N_i}(t_i)\ket{Vac} =\\ &
    \bra{Vac} a^\dagger_{N_i}(t_i) a_{N_i}(t_i) a^\dagger_{s}(t_{s}) a_{s}(t_{s})  \ket{Vac} \\
    =& N_i \bra{Vac}a^\dagger_{s}(t_{s}) a_{s}(t_{s})  \ket{Vac} = N_i R_s \frac{2}{\Gamma_{cav}}\kappa^2\\\\
    &\bra{Vac} a^\dagger_{N_s}(t_s)a^\dagger_{i}(t_{i}) a_{i}(t_{i})  a_{N_s}(t_s)\ket{Vac} =\\&
    \bra{Vac} a^\dagger_{N_s}(t_s) a_{N_s}(t_s) a^\dagger_{i}(t_{i}) a_{i}(t_{i})  \ket{Vac} \\
    =& N_s \bra{Vac}a^\dagger_{i}(t_{i}) a_{i}(t_{i})  \ket{Vac} = N_s R_i \frac{2}{\Gamma_{cav}}\kappa^2
    \end{aligned}
\end{equation}

Hence, the total coincidence rate is
\begin{equation}
\begin{aligned}
    G^{(2)}_{si}(t_s, t_i) =& R_i R_s[(\kappa^2 \times e^{-\Gamma_{cav} |t_s - t_i|}+ \\
    &\frac{4}{\Gamma_{cav}^2}\kappa^4)  + \frac{2}{\Gamma_{cav}}\kappa^2 (n_s + n_i) + n_i n_s]\\
    n_i \equiv & N_i / R_i\\
    n_s \equiv & N_s / R_s
\end{aligned}
\end{equation}

and 
\begin{equation}
    g^{(1)}_s = \frac{2}{\Gamma_{cav}} \kappa^2 R_s + N_s = R_s (\frac{2}{\Gamma_{cav}} \kappa^2 + n_s)
\end{equation}
\begin{equation}
    g^{(1)}_i = \frac{2}{\Gamma_{cav}} \kappa^2 R_i + N_i = R_i (\frac{2}{\Gamma_{cav}} \kappa^2 + n_i)
\end{equation}
and it turns out 
\begin{equation}
    \begin{aligned}
    g^{(2)}_{si}(t_s, t_i) = 1 + \frac{\kappa^2}{ (\frac{2}{\Gamma_{cav}} \kappa^2 + n_i)(\frac{2}{\Gamma_{cav}} \kappa^2 + n_s)}
    \\
    =1 + \frac{1}{(\frac{2}{\Gamma_{cav}} (n_s + n_i) + \frac{4}{\Gamma_{cav}^2} \kappa^2 + \frac{n_i n_s}{\kappa^2})}
    \end{aligned}
    \label{g2}
\end{equation}

\subsubsection{Best Coupling ($\kappa_{best}$) Point}
The optimized coupling,  $\kappa_{best}$, which maximizes $g^{(2)}_{si}$ in Eq.\ref{g2}, us. 
\begin{equation}
    \begin{aligned}
    \kappa^2_{best} = \frac{\Gamma_{cav}}{2}n\\
    n \equiv \sqrt{n_i n_s}
    \end{aligned}
\end{equation}
while the corresponding $g^{(2)}_{si}$ is
\begin{equation}
    \begin{aligned}
    g^{(2)}_{si}(0)|_{best} = 1 + \frac{\frac{\Gamma_{cav}}{2}}{(\sqrt{n_s} + \sqrt{n_i})^2}
    \end{aligned}
    \label{best-g2}
\end{equation}

In our experiment (Cs $D_l$-line quantum storage experiment), since (1). $N_s \gg N_i $ and (2). $R_s \ll R_i$, so $n_s >> n_i$. Thus, Eq.\ref{best-g2} yields,
\begin{equation}
    g^{(2)}_{si}(0)|_{best} \sim 1 + \frac{\Gamma_{cav}}{2n_s}
\end{equation}

\subsection{Characterization of noises\label{app: noises for g2}}
\begin{figure}[t!]
    \centering
    \includegraphics[width = 6cm]{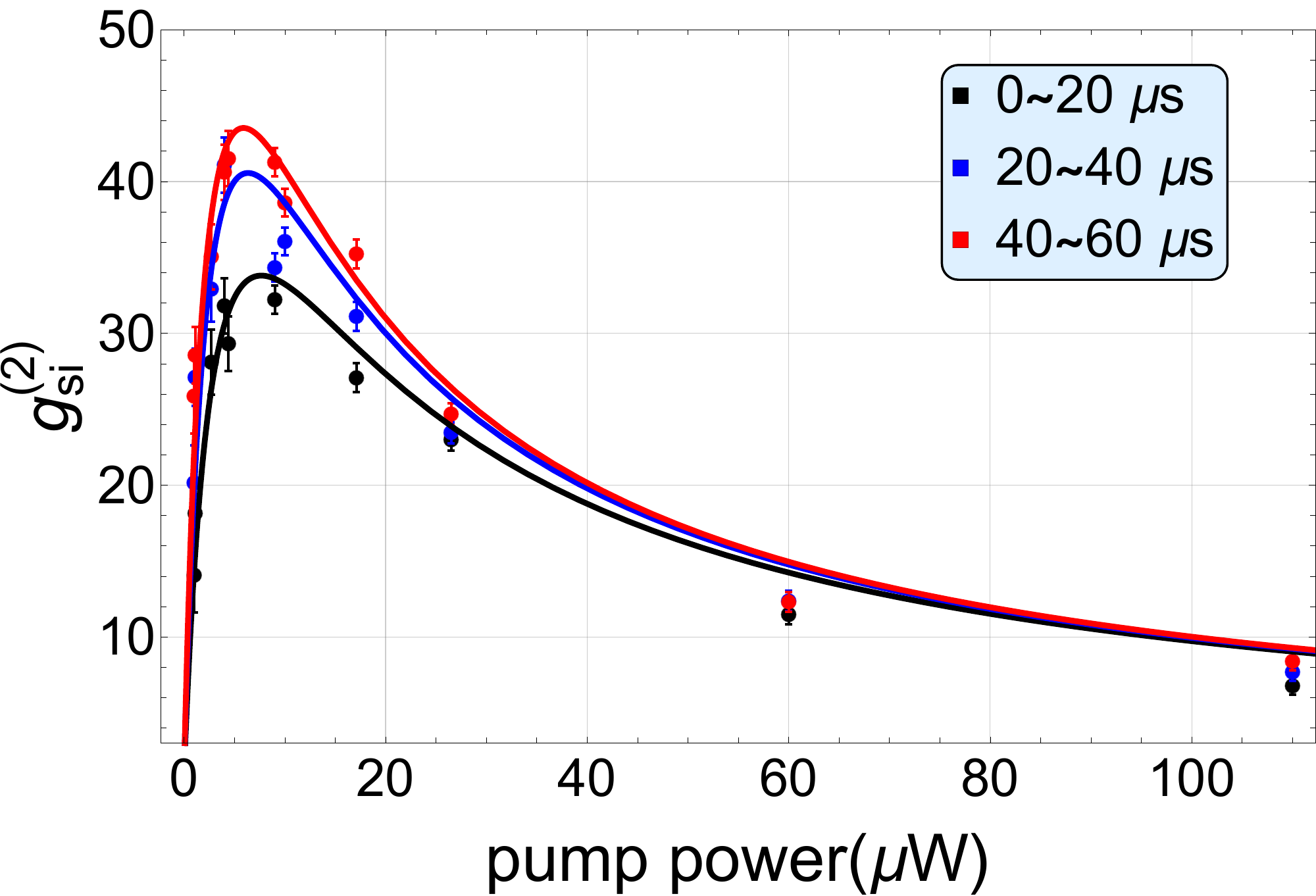}
    \caption{$g_{si}^{(2)}$ versus pump in $20 \mu sec$ time intervals with different delay time after the switching from OPO phase to SPDC phase. Note that all the time windows start after another $20 \mu s$ delay. The photon statistic result of earliest interval ($20\sim40 \mu s$) is degraded by the noise attributed to residue photons from OPO phase. The fitting parameters, $\{\Gamma, n_{s}, n_{i}, A\}$, for $\{0\sim20,  20\sim40, 40\sim60\}\mu s$ are $\{18., 0.059, 0.079,  0.08\}$, $\{18., 0.047, 0.067, 0.08\}$, and $\{18., 0.043, 0.063, 0.08\}$, respectively.}
    \label{res:cavnoise}
\end{figure}

The noises comprise environmental noises, intrinsic noises, and systematic noises. Environmental noises include dark counts ($\sim 100 \text{Hz}$) from the photon detectors and stray photons from the surrounding. To cope with them, thorough blockage of the whole system from the surrounding is necessary. Intrinsic noise results from the uncorrelated photons generated from the photon source (accidental counts) \cite{PhysRevA.83.061803}. We observe that the residue photons from the OPO phase contribute to the background of $g_{si}^{(2)}$ measurement in the SPDC phase. To investigate this problem, we split the detecting time interval ($60  \mu s$ shown in Fig.\ref{fig:setup}b) into three equal sub-interval and see the $g_{si}^{(2)}(0)$ of each at the photon source lab, as shown in Fig.\ref{res:cavnoise}. Note that another $20 \mu s$ delay has been added before each time interval. In the earliest interval ($0\sim20 \mu s$), residue photons remain in the cavity due to the high finesse. Consequently, the lowest $g_{si}^{(2)}(0)$ is presented at this interval. This can be referred to appendix A in \cite{Tsai_2018}.

\begin{figure}[t!]
  \centering
    \includegraphics[width = 6.5cm]{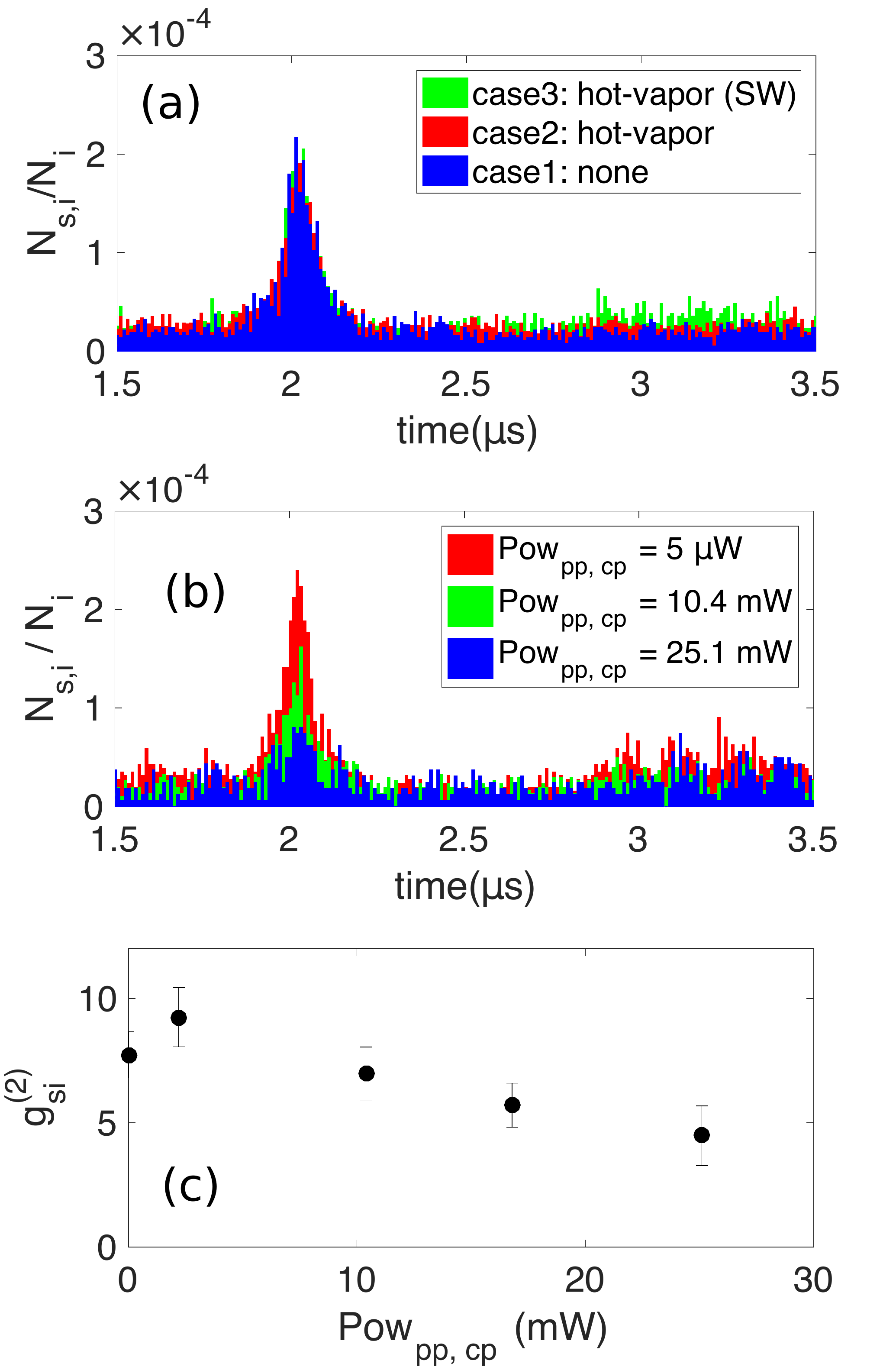}
    \caption{ (a). The measured coincidence counts normalized by the heralded idler counts. The measurements are conducted when the pumping control beam is not set, so that Raman-induced noise exists. The measurements are conducted when the 1. control beam is constantly on and the cell is without any hot vapor (blue-coloured area) 2. the control beam is constantly on for pumping and the hot vapor is filled with the cell  (red-coloured area), and 3. the control beam is switched off at the storage phase ($2.2\sim 2.8 \mu s$) and on at the reading stage (after $2.8 \mu s$) (green-coloured area), while the hot vapor is filled with the cell.  When the hot vapor is full of the chamber, it generates the Raman-induced noises, and thus the background level of $2, 3$ are higher. These additional noises are not from control beam's stray light, since the control beam is also on at the $1$ case. When using $3$, since the pumping is off at the storage, the accumulated $\ket{F=4}$ population increases. As a result, the noises increases when control beam is on. (b). The measured coincidence counts normalized by the heralded idler counts. When increasing the control beam power at the pumping stage (see Fig.\ref{fig:setup}b) the signal, the signal of singal photons is deteriorating. It is because the control beam is reflected back and enter into SPCM1 (see Fig.\ref{fig:setup}), which results in the increase of false triggers. $\text{Pow}_{pp, cp}$ denotes the pumping power of control beams. (c) Different $g^{(2)}_{si}$ with different pumping power of control beams.}
    \label{fig:suppl}
\end{figure}

Systematic noises consist of three main sources: stray noises from the control beam, Raman-induced noises, and false triggers on the idler detector from reflected control beam.  The first one can be filtered out by etalon filters (at the cost of total efficiency) and properly-selected control beam incident angle. Secondly, Raman-induced noises occure when at the reading stage, especially when using strong control field to read out. Even though we have prepared all atomic population at the $\ket{F=3}$ for EIT storage, some fraction of atomic population is back to $\ket{F=4}$ when waiting for the incoming signal (see Fig.\ref{fig:setup}a). Therefore, when the strong control beam ramps on at the reading stage, $\ket{F=4}$ population is pumped to the excited state ($\ket{F=4'}$) and decay into $\ket{F=3}$, thus generating unwanted Raman-induced noises. The phenomena are illustrated in Fig.\ref{fig:suppl}b, as we compare the data when the the cell is with/without hot vapors. To overcome this problem, we make the control beam constantly pumping the cold atomic ensemble until idler photons trigger the controlling switch. On the other hand, it is instructive to point out that the Raman-induced noises increase if the control beam is ramped off (storage phase) and ramped up (reading phase) again. It is because control beam serves as pumping, so if the control beam is off for a short time, the unwanted atomic population ($\ket{F=4}$) starts accumulating, as shown in Fig.\ref{fig:suppl}b. Conceivably, these Raman-induced noises become more rampant when storage time increases, since the $\ket{F=4}$ population of atoms can accumulate for longer time.

Thirdly, false triggers on the idler detector from reflected control beam also deteriorate the signal and are not so straightforward. A small fraction of strong control beam in the pumping stage (10 mW period in Fig.\ref{fig:setup}b) is reflected back by the etalon filter and, of minor portion, the surface of atomic vapor cell. Note that this pumping stage is to reduce the Raman-induced noises, as explained above. As shown in Fig.\ref{fig:suppl}a, the increasing control beam power leads to the declining of coincident counts. The systematic check is illustrated in Fig.\ref{fig:suppl}a.


\nocite{*}

\bibliography{apssamp}

\end{document}